\documentclass{elsart}

\usepackage{psfig}

\usepackage{natbib}

\newcommand{\et}{et al.\ }

\newcommand{\asca}{{\it ASCA}}
\newcommand{\xte}{{\it RXTE}}
\newcommand{\sax}{{\it BeppoSAX}}

\begin{document}

\runauthor{Cicero, Caesar and Vergil}


\begin{frontmatter}

\title{Evidence for a high accretion rate as the defining parameter of 
Narrow-Line Seyfert 1 galaxies} 

\author[XRA]{K. Pounds}
\author[XRA]{S. Vaughan}
\address[XRA]{X-Ray Astronomy Group; Department of Physics and
Astronomy; Leicester University; Leicester LE1 7RH; UK}

\begin{abstract}
X-ray spectral features which are unusually strong in many Narrow Line
Seyfert galaxies are found to be consistent with reflection from strongly
ionized matter, providing further evidence of a high accretion rate in
these objects and offering a unique signature of that key parameter in
future observations.
\end{abstract}


\end{frontmatter}


\section{Introduction}

The  abstracts  of papers  for  this  meeting  showed a  strengthening
consensus  that  the  defining  parameter  of Narrow  Line  Seyfert  1
galaxies (NLS1s) is  indeed a high accretion rate.   That proposal was
first made$^{1}$ by  direct analogy with the X-ray  properties of GBHC
in their  high state, where it  was also noted that the resulting
increase  in ionising  flux  would  lead to  the  `broad line'  clouds
forming  at larger  radii,  providing a  natural  explanation for  the
narrow permitted  optical lines characteristic of  NLS1.  A prediction
that the hard X-ray spectrum of  NLS1 would be unusually steep, due to
increased  cooling of  the Comptonising  electrons, was  borne  out by
subsequent \asca\ spectra$^{2}$.

In  the present  paper  we discuss  the  interpretation of  additional
spectral  features,  apparently characteristic  of  NLS1, which  lends
further  support to  the high  accretion  rate thesis  and offers  the
exciting potential of  a direct observational signature of  one of the
fundamental parameters of the AGN phenomenon.

\section{Reflection from an ionized disc}

The importance of `reflection' in modifying the observed X-ray spectra
of AGN,  in turn providing a  key diagnostic of  the accretion process
believed to be  the source of power in AGN,  was established almost 10
years ago$^{3}$.  For many Seyfert galaxies and low luminosity QSOs, a
good explanation  of the observed  hard X-ray spectral features  (Fe K
fluorescence line, 10--20~keV `hump')  is provided by re-processing of
order a half of the `primary' X-ray power law flux, in dense, cold matter
(the putative accretion disc).  A number of attempts have subsequently
been  made to model  the effects  of the  external irradiation  on the
ionisation  state of  the reflecting  material and  hence  predict the
resulting   modification  of   the  reflected   spectrum.    An  early
study$^{4}$ noted the potential significance of a high accretion rate,
and strong X-radiation, in ionising the outer disc layers, finding the
ionisation  parameter to  scale with  $\dot{m}^{3}$ (where  $\dot{m} =
\dot{M} / \dot{M}_{Edd}$).

\begin{figure}
\centerline{
\hbox{
\psfig{figure=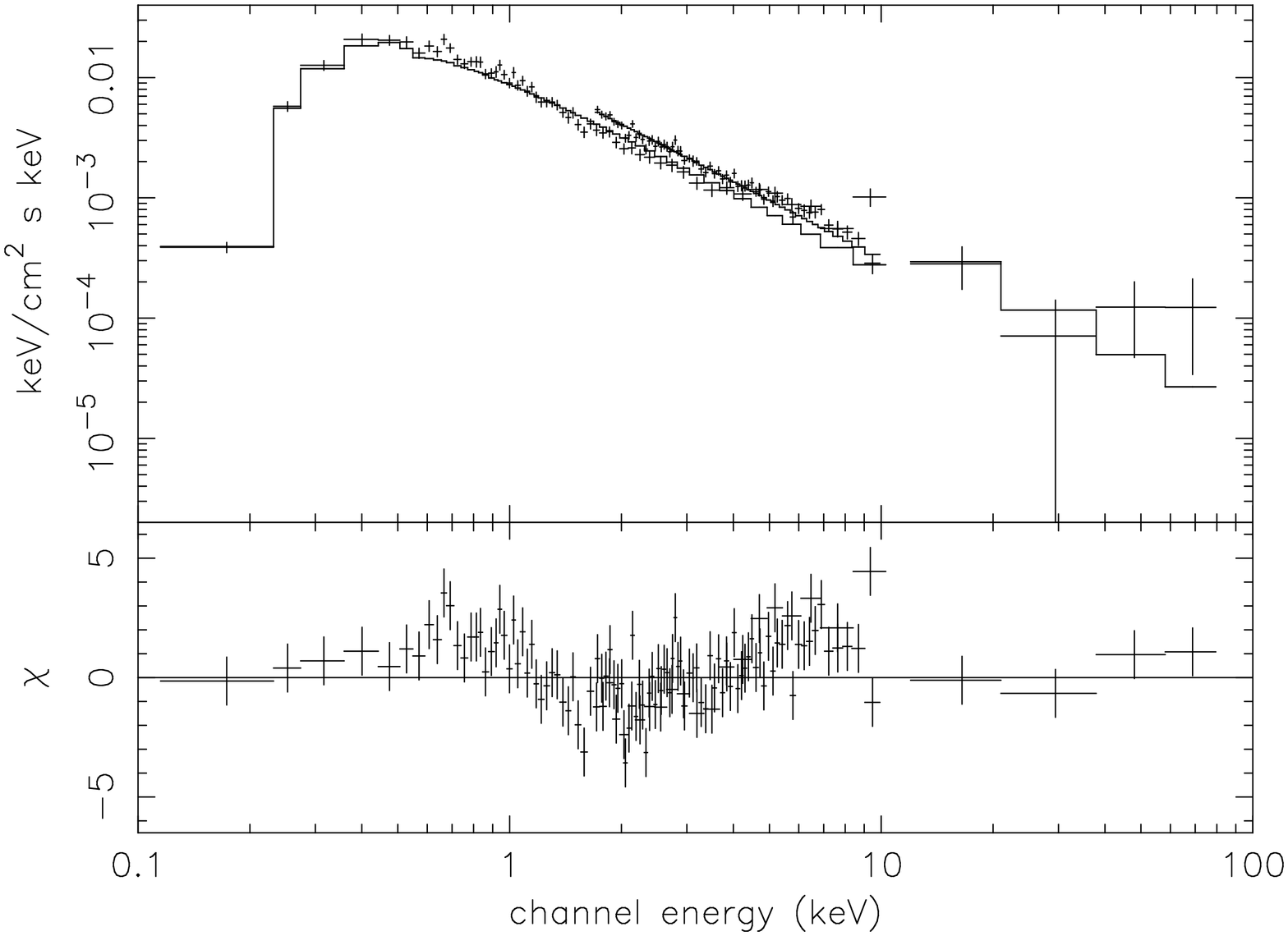,width=0.5\textwidth,height=0.255\textheight,angle=270}
\psfig{figure=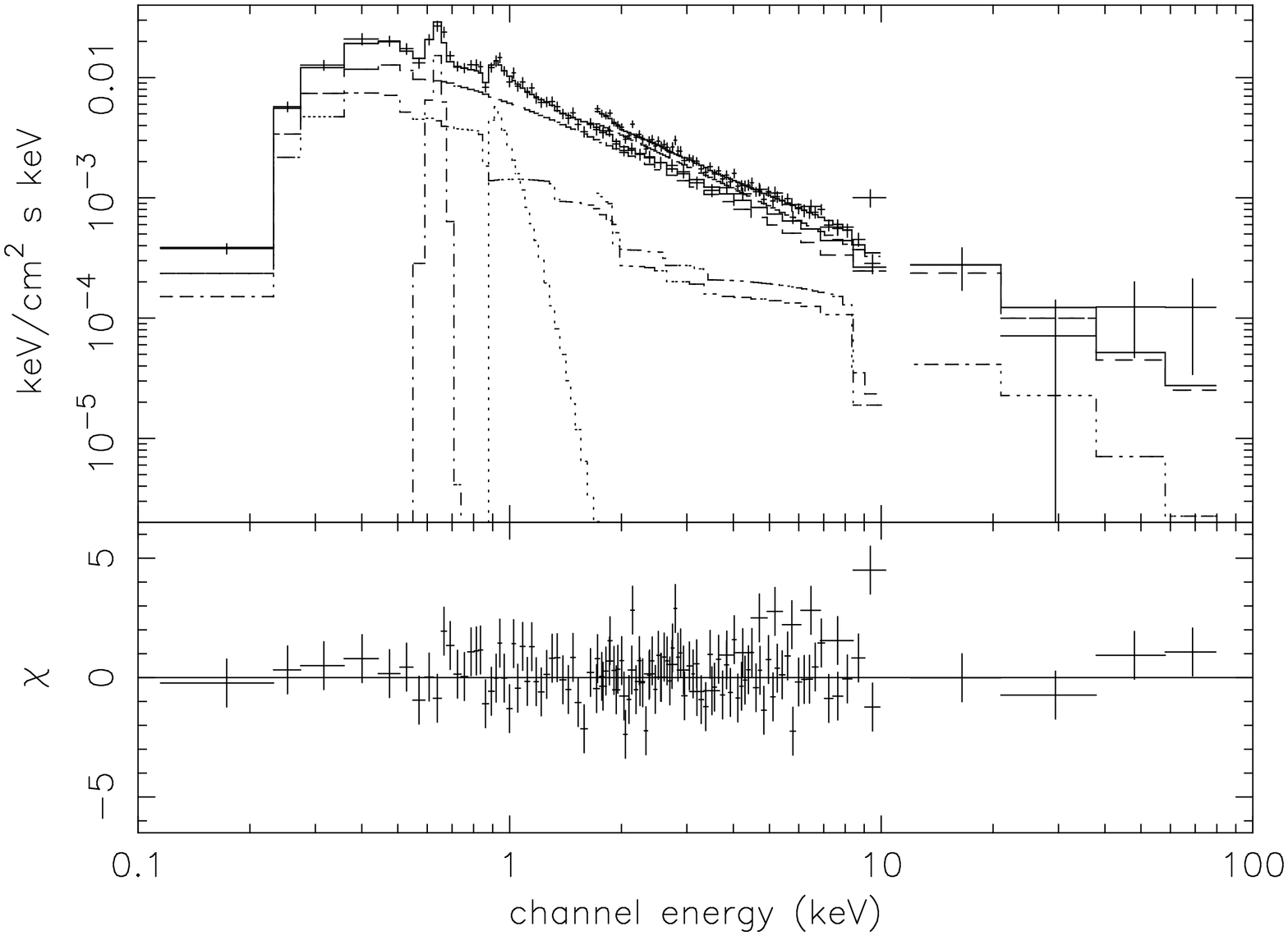,width=0.5\textwidth,height=0.255\textheight,angle=270}
}
}
\caption{(left)
Spectral fit to the \sax\ data of Ark 564 using a
simple power law model. Fig.~2. (right) shows a much improved fit using
the ionized disc model described in the text.
}
 \end{figure}

Until very recently, all ionized  disc models had made the simplifying
assumption of a constant density (Ross et al.$^{5}$ and refs therein),
notwithstanding that such a  situation is physically unrealistic.  One
constant density model (PEXRIV) has been adapted to the XSPEC spectral
fitting procedure, and has therefore  been used to assess recent X-ray
data from  \asca\ and  \sax. However, PEXRIV  also lacks  any emission
components.

A new approach has now been made to calculate the photoionisation, and
subsequent X-ray  reflection characteristics, of an  accretion disc in
which the density is determined  from a hydrostatic balance and solved
simultaneously   with    the   ionisation   balance    and   radiative
transfer$^{6}$.  The predictions of this, more physically
realistic model,  differ significantly  from  those  of the  constant
density models.  In particular, three discrete thermally stable layers
are predicted, whose optical  depths and effective temperatures depend
critically on the intensity {\it and} spectral index of the irradiating
X-rays.  In  the case  of a `normal'  Seyfert 1, the  outermost layer,
closest  to the  ionising  source,  is predicted  to  be almost  fully
ionized  and   at  the   local  Compton  temperature   ($\sim10^{7}  -
10^{8}$~K), while  at larger depths the  temperature decreases sharply
to  the thermal  disc temperature.   In terms  of the  reflected X-ray
spectrum, the  upper layer is  effectively a good  mirror, introducing
few spectral  features, while  the lowest layer  will act like  a cold
(neutral)  reflector, yielding the  Fe K  6.4~keV line  and 10--20~keV
reflection hump.

However,  with  a  steeper  (and  implicitly  more  intense)  incident
spectrum -- as is typical of  the ultrasoft NLS1s -- the Nayakshin \et
model predicts a region  of substantial optical depth with intermediate
temperature  (T$\sim 10^{6}$~K) and  ionisation. Reflection  from such
material  would  superimpose  significant  spectral  features  on  the
emerging X-ray spectrum, as is found with many NLS1.


A steep X-ray spectral slope is a primary feature of NLS1, of course.
Also, the interpretation that NLS1s operate at a high accretion rate
implies an unusually high intensity of soft XUV flux, precisely the
irradiation conditions required in the `hydrostatic balance' models$^{6}$
to give an intermediate temperature, stable region
with sufficient optical depth to imprint significant ionisation features
on the reflected X-ray spectrum.

A recent study of the bright NLS1 Ark 564$^{7}$, in which
\asca\ and \xte\ data have been combined to give simultaneous spectral
coverage from 1--20~keV, has revealed several features
identified as arising from reflection in an ionized disc. See also Vaughan
\et in these Proceedings.

\section{A new \sax\ spectrum of Ark 564}

While  our \asca/\xte\  study of  Ark~564 provides  good  evidence for
ionized reflection features, in terms  of the Fe K-absorption edge and
`soft excess' near 1~keV, the instrumental uncertainties of the \asca\
data restricted  the analysis below  $\sim 1$~keV.  We  have therefore
tested  our spectral fitting  against archived  \sax\ data,  where the
combination of the  LECS, MECS and PDS instruments  provide good cover
down to $\sim 0.2$~keV (and to above 15~keV).  Figure 1 shows a simple
power law is an unacceptable  fit to the \sax\ data ($\chi^{2}_{\nu} =
1.34$), with  residual features similar to those  from the \asca/\xte\
fit. Figure  2 reproduces  the fit  to a power  law plus  ionized disc
reflector, the latter being  modelled by PEXRIV plus emission features
of O~\textsc{viii} Lyman-$\alpha$  at 0.65~keV and the O~\textsc{viii}
recombination continuum above 0.87~keV.  The equivalent widths of both
oxygen  emission features  are $\sim$90~eV  and the  excellent overall
spectral fit has a $\chi^{2}_{\nu}$ of 1.02. We draw particular attention
to  the strength  of  the O~\textsc{viii}  recombination continuum,  a
feature largely  ignored in previous  attempts to fit  the $\sim$1~keV
emission  features  in NLS1  (but  included  in  the spectral  fit  of
Mrk~3$^{8}$).  The O~\textsc{viii}  recombination continuum also shows
up strongly in  the relevant models of Nayakshin  et al.$^{6}$, and we
further note the self-consistency of the ionized disc explanation with
a temperature  of $\sim 10^{6}$~K  derived from the measured  width of
the recombination continuum.

\section{Discussion}

There  is now good  evidence that  another distinguishing  property of
NLS1 is the  presence of strong spectral features  supersimposed on an
intrinsically steep power law. At least  for Ark 564, we find that ALL
the observed spectral features,  from the ionized Fe K-absorption edge
to the `soft excess' below  $\sim1$~keV, can be interpreted as arising
from  an  ionized  disc.   We  note  encouraging  agreement  with  the
predictions of the model of Nayakshin \et, where the unusual intensity
and steepness  of the intrinsic X-ray  emission of a  NLS1 is critical
for a  significant optical depth of intermediate  ionisation to occur.
We  note, in  passing, that  this interpretation  offers  the exciting
future prospect of  using X-ray spectra as a  unique diagnostic of the
accretion rate in AGN.

Finally, a comment  on the question of NLS1 being  `low mass' or `high
accretion  rate' objects,  a debate  continued in  this  meeting. That
difference may  be merely semantic,  given that the NLS1  and `normal'
Seyfert 1 galaxies accessible to  current observation lie in a limited
range  of luminosity.   Correspondingly,  NLS1 tend  to  be both  high
accretion  rate and low  mass (for  their luminosity).  More sensitive
future  X-ray observations  should allow  the expected  differences in
appearance  of high  mass AGN  with  different accretion  rates to  be
studied.



\begin{thebibliography}{8}
\bibitem{1} Pounds, K., Done, C., Osborne, J.\ 1995, MNRAS, 277, L5
\bibitem{2} Brandt, W. N., Mathur, S., Elvis, M.\ 1997, MNRAS, 285, L25
\bibitem{3} Pounds K. A. \et 1990, Nature 344 132
\bibitem{4} Ross R. R., Fabian A. C.\ 1993, MNRAS 261 74
\bibitem{5} Ross R. R., Fabian A. C., Young A. J.\ 1999, MNRAS 306 461
\bibitem{6} Nayakshin S., Kazanas D., Kallman T. R.\ 2000, ApJ, in
press 
\bibitem{7} Vaughan, S. \et 1999, MNRAS, 308, L34
\bibitem{8} Griffiths R. G. \et 1998, MNRAS 298 1159 
\end{thebibliography}
\end{document}